\documentclass[a4paper,fleqn,usenatbib]{mnras}
\usepackage{newtxtext,newtxmath}

\usepackage[T1]{fontenc}
\usepackage{ae,aecompl}

\usepackage{graphicx}	
\usepackage{amsmath}	
\usepackage{amssymb}	

\title[Line-driven disc wind in near-Eddington AGNs]{Line-driven disc wind in near-Eddington active galactic nuclei: decrease of mass accretion rate due to powerful outflow}

\author[M. Nomura et al.]{
Mariko Nomura,$^{1,2,3}$\thanks{E-mail: m-nomura@kure-nct.ac.jp}
Ken Ohsuga,$^{4,5,6}$
and Chris Done$^{7,8}$
\\
$^{1}$Faculty of Natural Sciences, National Institute of Technology (KOSEN), Kure
College, 2-2-11 Agaminami, Kure, Hiroshima 737-8506, Japan\\
$^{2}$Astronomical Institute, Tohoku University, 6-3 Aoba, Aramaki, Aoba-ku, Sendai, Miyagi, 980-8578, Japan\\
$^{3}$Department of Physics, Faculty of Science and Technology, Keio University, 3-14-1 Hiyoshi, Kohoku-ku, Yokohama, Kanagawa 223-8522, Japan\\
$^{4}$Center for Computational Sciences, University of Tsukuba, Ten-nodai, 1-1-1 Tsukuba, Ibaraki 305-8577, Japan\\
$^{5}$National Astronomical Observatory of Japan, Osawa, Mitaka, Tokyo 181-8588, Japan\\
$^{6}$School of Physical Sciences, Graduate University of Advanced Study (SOKENDAI), Shonan Village, Hayama, Kanagawa 240-0193, Japan\\
$^{7}$Centre for Extragalactic Astronomy, Department of Physics, University of Durham, South Road, Durham DH1 3LE, UK\\
$^{8}$Visiting senior scientist, Kavli IPMU,The University of Tokyo, 5-1-5 Kashiwanoha Kashiwa, 277-8583, Japan
}

\date{Accepted 2020 April 1. Received 2020 March 31; in original form 2018 November 1}

\pubyear{2020}

\begin{document}
\label{firstpage}
\pagerange{\pageref{firstpage}--\pageref{lastpage}}
\maketitle

\begin{abstract}
  Based on recent X-ray observations, 
  ultra-fast outflows from supermassive black holes are expected to
  have enough energy to dramatically affect their host galaxy 
  but their
  launch and acceleration mechanisms are not well understood.
  We perform two-dimensional radiation hydrodynamics simulations of UV
  line-driven disc winds in order to calculate the mass loss rates and
  kinetic power in these models. We develop a new iterative technique
  which 
  reduces the mass accretion rate through the
  inner disc in response to the wind mass loss. This makes the inner
  disc is less UV bright, reducing the wind power compared to previous
  simulations 
  which assumed 
  a constant accretion rate with radius.
  The line-driven winds in our simulations are still extremely
  powerful, with around half the supplied mass accretion rate being
  ejected in the wind for black holes with mass $10^8$--$10^{10}\,M_\odot$
  accreting at $L/L_{\rm Edd}=0.5$--$0.9$. 
  Our results open up the way for estimating the growth rate 
  of supermassive black hole and evaluating the kinetic energy ejected 
  into the inter stellar medium (active galactic nuclei feedback) 
  based on a physical model of line-driven disc winds.
\end{abstract}




\section{Introduction} \label{sec:intro}
Supermassive black holes (SMBHs) are found in the centre of almost all
known galaxies, and their masses are observed to correlate with the 
mass of stars in their host galaxy bulge. This implies that the star
formation powered growth of the galaxy is linked to the accretion
powered growth of its black hole via feedback
\citep[e.g.,][]{1998AJ....115.2285M,2000ApJ...539L...9F,2000ApJ...539L..13G,2002ApJ...574..740T}. 
The details of this feedback mechanism are not well understood, 
but winds powered by the accretion flow onto the black hole
are likely to play a role. 
This is supported by recent observations of 
ultra-fast outflows (UFOs) in active galactic nuclei (AGNs). These
winds are identified via blueshifted absorption
lines of highly ionized iron (FeX\hspace{-.1em}X\hspace{-.1em}V and/or
FeX\hspace{-.1em}X\hspace{-.1em}V\hspace{-.1em}I) detected in the
X-ray band. 
These were first detected in a few bright quasars 
\citep[e.g.,][]{2002ApJ...579..169C,Pounds03,2003ApJ...593L..65R}, but more recent systematic studies
of archival data show that these are likely a common feature in AGN
\citep{Tombesi10,Tombesi11,Tombesi12a,2013MNRAS.430...60G,Gofford15}. 
The typical outflow
velocity is $0.1$--$0.3c$, where $c$ is the speed of light, giving an
estimate for the kinetic power of (0.1--10\%)$L_{\rm  Edd}$, where $L_{\rm Edd}$ is the Eddington luminosity 
\citep{Tombesi12a,Gofford15}.  
This is large enough to affect the properties of the host galaxy, possibly setting the $M$--$\sigma$ relation \citep{2003ApJ...596L..27K}.

The terminal velocity of a wind typically is of order the escape
velocity from its launch point, so the fast velocity of the UFOs means
that they should be launched from the accretion disc close to the
SMBH. There are only a few potential mechanisms to drive a wind from
such high gravity regions: radiation pressure on electrons for
super-Eddington sources (super-Eddington winds), radiation pressure on UV
line transitions (UV line-driven winds), and magnetic fields (magnetic
winds). Most UFOs are seen in sources which are not above the
Eddington limit, leaving only UV line driving or magnetic winds. UV
line driving requires that the material has low ionization state,
where radiation pressure on the multiple UV spectral lines produces a
force, pushing material away from the SMBH.
Theoretical studies of UV line-driven winds started from an analytic approach
\citep{1995ApJ...451..498M}
but now include both hydrodynamic simulations
\citep[][hereafter PK04]{Proga00,Proga04}
and non-hydrodynamic calculations of streamlines from ballistic trajectories
\citep{Risaliti10,Nomura13}.
The line force is $10$--$1000$ times larger than that due to radiation on electrons when
the material is in a low-ionization state \citep{SK90}, leading to a
high-velocity disc wind. However, the observed very high
ionization state of the UFOs is apparently in conflict with UV line
driving \citep{Tombesi10}, but this mechanism can still work if the
wind is only highly ionized after it has reached terminal velocity
(e.g., \citeauthor{2015MNRAS.446..663H} \citeyear{2015MNRAS.446..663H}; Mizumoto et al. 2020 in preparation)

Simulations of UV line-driven winds show that UV line driving can
indeed produce powerful winds from sub-Eddington AGN despite X-ray irradiation
(\citealt{Proga00}; PK04; \citealt{Risaliti10,Nomura13}).
These simulations include the effect of a central X-ray source in ionizing the wind, and
typically show that a UV line-driven wind is launched from the UV
bright disc, lifting material up to where it intercepts more of the
hard X-ray coronal radiation, so is overionized. This removes the UV
transitions, so the line force drops and the outflow stalls and
falls back, producing a vertical structure which shields material
further out from the ionizing radiation. The failed wind region
becomes larger and larger until the shielding is sufficient for
material in the outer regions to reach its escape velocity before it
is overionized by the X-ray flux.

These hydrodynamic models predict the density, ionization state and velocity of material 
around the disc.
However, 
the absorption features in the outflow are
due to multiple line and bound-free transitions, so a full prediction of the
spectral features for detailed comparison to observations 
requires postprocessing of the results using a full photo-ionization 
radiation transfer code. This is a very intensive calculation so has only been performed 
for a single hydrodynamics simulation, that of PK04 for a black hole mass of 
$M_{\rm BH}=10^8
\,M_{\odot}$ and 
luminosity of
$L/L_{\rm Edd}=0.5$ for a central X-ray source with
$L_{\rm X}=0.1L_{\rm D}$,
where $L_{\rm D}$ is the luminosity of the accretion disc.
This set of parameters resulted in a
funnel-shaped wind with high ionization and 
high velocity of $\sim 0.05$--$0.1c$, carrying away
around half of the total mass accretion rate though the disc.
This can produce blueshifted absorption lines qualitatively similar to 
those seen in the UFOs \citep{Schurch09,Sim10,HP14}, though this
particular simulation does not produce high enough wind velocity 
to fit the fastest UFOs observed 
\citep[e.g., PDS456,][]{Reeves09}.
Instead, magnetic winds can 
easily explain the high-velocity, high ionization 
outflowing matter \citep[e.g,][]{B82,K94,Fukumura15}, 
but this model requires that the magnetic field lines form a 
specific global topology \citep{B82} 
which cannot (currently) be determined {\it ab initio}. 

Instead, our previous works \citet{N16} and \citet[][hereafter
N17]{N17} calculated 
radiation hydrodynamics simulations of UV
line-driven winds over a much wider parameter space of black hole mass, mass
accretion rate, and X-ray irradiation. These can be used to estimate the absorption column 
along each line of sight, velocity and typical ionization state of the material for a rough 
comparison with the data. The results of N17 show that these UV line-driven wind hydrodynamics simulations
do indeed span the range observed in UFOs, and also
can match the luminosity dependence of the mass outflow rate
suggested by \citet{Gofford15}. 
These results clearly show that UV line-driven winds are energetically consistent with the
observed UFOs. 

However, all such simulations to date are not self-consistent as 
they assume a 
constant accretion rate with radius.
Yet many of the predicted
winds, especially those at $L/L_{\rm Edd}\gtrsim 0.5$, have mass loss
rate, $\dot{M}_{\rm out}$, which is a large fraction of the mass
accretion rate through the disc.
This is important
as the most powerful UFOs observed are in sources with similarly large
$L/L_{\rm Edd}$, where the winds should expel 
large amounts of mass from the accretion disc. 
 
In this paper, we improve the 
calculation method of the radiative
hydrodynamic code of N17
so as to consider
the reduction of mass accretion rate due to the 
loss of mass and angular momentum in the UV line-driven disc
wind.
This reduces the UV flux from the inner disc, but powerful UV
winds are still produced, and carry enough energy and momentum
to impact on the host galaxy growth.

\section{Method}\label{sec:methods}
\subsection{Basic equations}
\label{sec:basic_eq}
We modify the calculation method of N17 
so as to satisfy the conservation law
of the total mass of the disc and winds.
We briefly review our calculation method, and then 
describe how we incorporate the reduction in mass accretion rate through the inner disc
caused by the wind losses.  

The simulations use spherical
polar coordinate $(r,\theta,\varphi)$, where $r$ is the distance 
from the origin of the coordinate, 
$\theta$ is the polar angle, and $\varphi$ is the azimuthal angle. 
We perform the calculation in  two-dimensions as we 
assume the axial symmetry about the rotation axis of the disc.
The basic equations of the hydrodynamics are
the equation of continuity,
\begin{equation}
  \frac{\partial \rho}{\partial t}+\nabla\cdot(\rho \mbox{\boldmath $v$})
  =0,
  \label{eoc}
\end{equation}
the equations of motion,
\begin{equation}
  \frac{\partial (\rho v_r)}{\partial t}+\nabla\cdot(\rho v_r \mbox{\boldmath $v$})
  =-\frac{\partial p}{\partial r}+\rho\Bigg[\frac{v_\theta^2}{r}+\frac{v_\varphi^2}{r}+g_r+f_{{\rm rad},\,r}\Bigg],
  \label{eom1}
\end{equation}
\begin{equation}
  \frac{\partial (\rho v_\theta)}{\partial t}+\nabla\cdot(\rho v_\theta \mbox{\boldmath $v$})
  =-\frac{1}{r}\frac{\partial p}{\partial \theta}+\rho\Bigg[-\frac{v_r v_\theta}{r}+\frac{v_\varphi^2}{r}\cot \theta+g_\theta+f_{{\rm rad},\,\theta}\Bigg],
  \label{eom2}
\end{equation}
\begin{equation}
  \frac{\partial (\rho v_\varphi)}{\partial t}+\nabla\cdot(\rho v_\varphi \mbox{\boldmath $v$})
  =-\rho\Bigg[\frac{v_\varphi v_r}{r}+\frac{v_\varphi v_\theta}{r}\cot \theta\Bigg],
  \label{eom3}
\end{equation}
and the energy equation,
\begin{equation}
  \frac{\partial}{\partial t}\Bigg[  \rho \Bigg(\frac{1}{2}v^2+e\Bigg) \Bigg]
  +\nabla \cdot \Bigg[  \rho\mbox{\boldmath $v$} \Bigg(\frac{1}{2}v^2+e+\frac{p}{\rho}\Bigg) \Bigg]
  =\rho\mbox{\boldmath $v$}\cdot\mbox{\boldmath $g$}+\rho\mathcal L,
  \label{eoe}
\end{equation} 
where $\rho$ is the mass density,
\mbox{\boldmath $v$}$=(v_r,\,v_\theta,\,v_\varphi)$
are the velocities,
$p$ is the gas pressure, $e$ is the internal energy per unit mass,
\mbox{\boldmath $g$}$=(g_r,\,g_\theta )$ 
is the gravitational acceleration of the black hole located at $(r,\theta)=(z_0,\pi)$, 
and \mbox{\boldmath $f$}$_{\rm rad}=(f_{{\rm rad},r},\,f_{{\rm rad},\theta})$ is the radiation force per unit mass (see Section \ref{sec:line} for detail).
An adiabatic equation of state $p/\rho=(\gamma -1)e$ with $\gamma=5/3$ 
is assumed in our calculations.

In the last term of Eq. \ref{eoe}, 
$\mathcal L$ is the approximate formula of the net heating rate introduced by \citet{1994ApJ...435..756B},
\begin{equation}
  \mathcal L=n^2(G_{\rm Compton}+G_{\rm X}-L_{\rm b,l}),
\end{equation}
where, $n$ is the number density, $G_{\rm{Compton}}$ is the heating rate
via the Compton scattering,
\begin{eqnarray}
  G_{\rm{Compton}}=8.9\times10^{-36}\xi(T_{\rm X}-4T)\, {\rm erg\,cm^{-3}\,s^{-1}},
\end{eqnarray}
$G_{\rm X}$ is the difference between the heating rate by 
the X-ray photoionization and the cooling rate via the recombination,
\begin{eqnarray}
  G_{\rm X}=1.5\times10^{-21}\xi^{1/4}T^{-1/2}(1-T/T_{\rm X}) \, {\rm erg\,cm^{-3}\,s^{-1}},
\end{eqnarray}
and $L_{\rm{b,l}}$ is the cooling rate by the bremsstrahlung and line emission,
\begin{eqnarray}
  \nonumber L_{\rm{b,l}}=3.3\times10^{-27}T^{1/2}+1.7\times10^{-18}\xi^{-1}T^{-1/2}\\
  \times \exp(-1.3\times10^5/T)+10^{-24} \, {\rm erg\,cm^{-3}\,s^{-1}}.
  \label{eq:bl}
\end{eqnarray}
In Eq.\ref{eq:bl}, the first term represents the effect of the bremsstrahlung and the last two terms show the line coolong in an optically thin atmosphere.
The temperature $T$ is obtained by
$T=\mu m_{\rm p} p/k_{\rm B}\rho$, 
where $\mu$ (=0.5) is the mean molecular weight,
$m_{\rm p}$ is the proton mass,
and $k_{\rm B}$ is the Boltzmann constant.
The ionization parameter, $\xi$, is 
defined as 
\begin{equation} 
  \xi=\frac{4\pi F_{\rm X}}{n}\, {\rm erg \,s^{-1}\,cm},
  \label{xi}
\end{equation}
where $F_{\rm X}$ is the X-ray flux.
In the present study,
the X-ray is assumed to come from the point source
at the centre
with luminosity $L_{\rm X}=f_{\rm X} L_{\rm D}$,
where $f_{\rm X}$ is a parameter.
We also assume the temperature of the X-ray radiation to be
$T_{\rm X} =10^8\,\rm{K}$.
The X-ray flux is attenuated
via the absorption of which the cross section 
is set to be $\sigma_{\rm X}=\sigma_{\rm e}$ for $\xi> 10^5$, or $100\sigma_{\rm e}$ for $\xi\le 10^5$,
where $\sigma_{\rm e}$ is the mass-scattering coefficient for free electrons.
Such a treatment is employed in previous works
\citep[][N17]{Proga00, N16}.
We note that the scattered and reprocessed photons are ignored in our simulations (see also Section \ref{sec:dis2}).

\subsection{Line force}
\label{sec:line}
The radiation force in Eq.\ref{eom1} and Eq.\ref{eom2} is described as 
\begin{equation}
  { \mbox{\boldmath $f$}_{\rm rad}}=\frac{\sigma_{\rm e} \mbox{\boldmath $F$}_{\rm D}}{c}+\frac{\sigma_{\rm e} \mbox{\boldmath $F$}_{\rm line}}{c}M,
  \label{radforce}
\end{equation}
where 
\mbox{\boldmath $F$}$_{\rm D}$ is the radiation flux of the disc emission 
integrated by the wavelength throughout the entire range, 
and \mbox{\boldmath $F$}$_{\rm line}$ is the 
line-driving flux (UV flux),
which is the same as \mbox{\boldmath $F$}$_{\rm D}$ but integrated across the UV transition band of $200$--$3200\,$\AA.
Both fluxes are calculated by integrating the radiation
from the inner hot region of the disc,
where the effective temperature is larger than $3\times 10^3\,{\rm K}$. 
We divide the disc surface into 
$4096$ grids for radial direction and
$4096$ grids for azimuthal direction,
and numerically calculate
$\mbox{\boldmath $F$}_{\rm D}$ and $\mbox{\boldmath $F$}_{\rm line}$.
The radial components of these fluxes are
attenuated due to electron scattering,
of which the optical depth is measured from
the origin of the coordinate.
On the other hand, $\theta$-components of 
$\mbox{\boldmath $F$}_{\rm D}$ and $\mbox{\boldmath $F$}_{\rm line}$
are supposed not to be diluted.

The force multiplier, $M$, is defined by \citet{SK90}, 
\begin{equation}
  M(t,\xi)=kt^{-0.6}\Biggl[ \frac{(1+t\eta_{\rm{max}})^{0.4}-1}{(t\eta_{\rm{max}})^{0.4}}\Biggr].
  \label{forceM}
\end{equation}
Here, $t$ is the local optical depth parameter written as,
\begin{equation}
  t=\sigma_{\rm e} \rho v_{\rm{th}}\Bigl| \frac{dv}{ds}\Bigr|^{-1}.
  \label{t-xi}
\end{equation}
Here, the thermal speed, $v_{\rm th}$,
is set to $20 \, {\rm km \,s^{-1}}$
that corresponds to the thermal speed
of the hydrogen gas of the temperature of $25,000\, {\rm K}$.
The velocity gradient, $dv/ds$, depends on the direction of each light-ray from the disc surface to the point of interest. However, in order to reduce the computational cost, we evaluate $dv/ds$ along the direction of the disc line-driving flux $($\mbox{\boldmath $F$}$_{\rm line}/|$\mbox{\boldmath $F$}$_{\rm line}|)$ in the same manner as N17.
This is because the radiation along this direction
is thought to mainly contribute the line force.
We note that this assumption gives the line force comparable to that evaluated by employing more accurate method (see Appendix \ref{app:FM}).

In Eq.\ref{forceM}, $k$ and $\eta_{\rm{max}}$ are the functions of the ionization parameter $\xi$ as,
\begin{equation}
  k=0.03+0.385\exp(-1.4\xi^{0.6}),
\end{equation}
and 
\begin{equation}
  \log_{10}\eta_{\rm{max}}=\left\{ 
  \begin{array}{ll}
    6.9\exp(0.16\xi ^{0.4}) & \log\xi \le 0.5 \\
    9.1\exp(-7.96\times 10^{-3}\xi) & \log\xi >0.5 \\
  \end{array} \right.
  .
  \label{k_eta}
\end{equation}

\subsection{Iterative method}
The main difference in computational method is 
the treatment of the mass accretion rate.
In N17, we assumed the mass accretion rate 
was constant with radius, as in the standard disc assumptions \citep{SS73}. This is
valid when the 
mass outflow rate via the disc wind is relatively small
in comparison with the mass accretion rate.
However, such a treatment breaks down for the most powerful winds.  

Here we extend the code to take into account the 
reduction in mass accretion rate through the disc caused by the mass
outflow rate. We approximate the total mass loss rate $\dot{M}_{\rm out}$ as coming from
a single radius, $R_{\rm launch}$. Thus the 
mass accretion rate for $R\ge R_{\rm launch}$ is the original 
mass accretion rate supplied through the outer disc, $\dot{M}_{\rm sup}$,
but it drops to $\dot M_{\rm BH}=\dot{M}_{\rm sup} -\dot{M}_{\rm out}$ for $R<R_{\rm launch}$. 

We estimate $R_{\rm launch}$ by conserving angular momentum per unit mass as well as mass.
The material lost in the wind retains the angular
momentum produced by the Keplerian velocity at the disc surface. 
Thus $R_{\rm launch}=l^2/GM_{\rm BH}$, 
where $l=\dot l_{\rm out}/\dot M_{\rm out}$ is the 
specific angular momentum of the wind material, calculated via
integrating over all angles
\begin{equation}
  \dot l_{\rm out}=
  4\pi r^2 \int^{89^\circ}_0 \rho r v_\varphi \sin ^2 \theta d\theta,
\end{equation} 
and
\begin{equation}
  \dot M_{\rm out}=
  4\pi r^2 \int^{89^\circ}_0 \rho v_r \sin \theta d\theta,
\end{equation} 
at the outer boundary.

Our iterative method consists of following procedures.
\begin{enumerate}
\item 
  Based on the standard disc model,
  of which the mass accretion rate is
  $\dot M_{\rm BH} (=\dot M_{\rm sup}-\dot M_{\rm out})$ 
  for the region of $r< R_{\rm launch}$
  and $\dot M_{\rm sup}$
  for the region of $r\geq R_{\rm launch}$,
  we calculate the radiation fluxes,
  $F_{\rm D}$ and $F_{\rm line}$,
  and set the temperature and the density on the $\theta=\pi/2$
  plane (see Section \ref{sec:boundary} for detail).
  Here note that, for the first iteration step,
  $\dot M_{\rm BH}$ and $R_{\rm launch}$
  can be set arbitrary within the range of
  $0 < \dot M_{\rm BH}\leq \dot M_{\rm sup}$ and $r_{\rm i}\leq R_{\rm launch} <r_{\rm o}$,
  here $r_{\rm i}$ and $r_{\rm o}$ are the radii
  at the inner and outer boundaries of the computational box.
  
\item
  We perform the hydrodynamics simulations with using above setup
  and estimate the time averaged mass outflow rate,
  $\langle \dot M_{\rm out} \rangle$,
  every $2\times 10^4 R_{\rm S}/c$,
  where  $R_{\rm S}$ is the Schwarzschild radius.
  We continue the hydrodynamics simulations
  until the time averaged mass outflow rate
  is converged within 10\%.
  
\item 
  We update the mass outflow rate with using 
  the convergence value of the time averaged mass outflow rate,
  $\dot M_{\rm out}=\langle \dot M_{\rm out} \rangle$.
  We also update $R_{\rm launch}$ 
  using the updated $\dot M_{\rm out}$. 
  
\item If either updated $\dot{M}_{\rm out}$ or $R_{\rm launch}$
  deviates from the value of the previous iteration step more than 10\%,
  we return back to the procedure (i). 
  We iterate procedures (i--iii)
  until $\dot M_{\rm out}$ and $R_{\rm launch}$ are
  converged within 10\%. 
\end{enumerate}

The number of the iterations to obtain the quasi-steady state
by above procedures depends on the parameters
($M_{\rm BH}$ and $\dot m_{\rm sup}=\dot M_{\rm sup}/\dot M_{\rm Edd}$)
and initial choice of $\dot M_{\rm BH}$ and $R_{\rm launch}$.
Here, $\dot{M}_{\rm Edd}$ is defined as $\dot{M}_{\rm Edd}=L_{\rm
Edd}/\eta c^2$ with the energy conversion efficiency $\eta=0.06$. 
For example, we need 12 times of iterations for 
$M_{\rm BH}=10^8\,M_\odot$
and $\dot m_{\rm sup}=0.5$
when we initially employ $\dot M_{\rm BH}$=$\dot M_{\rm sup}$
(then, we do not need to set $R_{\rm launch}$ because of
$\dot{M}_{\rm out}=0$).

As we have mentioned above, in our method,
we simply assume that the mass accretion rate
discontinuously changes at the radius of $r=R_{\rm launch}$.
However, the mass accretion rate would gradually decrease
in the wind launching region practically.
Also, we assume that the disc is stable and steady.
The density and the temperature at the $\theta=\pi/2$ plane
are calculated by assuming 
that the disc is in the hydrostatic equilibrium in the vertical direction.
However, \citet{1986ApJ...306...90S} calculated the time dependent disc structure 
and found that the disc can exhibit the periodic variation
of the accretion rate when the mass loss rate is sufficiently large. 
It is necessary to investigate
the time variation of the disc structure 
via the launching the line-driven winds
in order to establish a more realistic wind model.
Such a study is left as an important future work.

\subsection{Boundary and initial conditions}
\label{sec:boundary}

The hydrodynamics are calculated over a computational domain of $r_{\rm i} =30R_{\rm S}\leq r \leq r_{\rm o}=1500R_{\rm S}$ and $0\leq \theta\leq \pi/2$.
The coordinate system is
set so that the black hole is located at $r=z_0$ and $\theta=\pi$, i.e., at cylindrical radius of 0, but a 
distance $z_0$ below the origin of the coordinate, where $z_0=3.1 \dot{m}_{\rm sup} R_{\rm S}$ which is the 
thickness of a standard Shakura-Sunyaev disc at $r_{\rm i}=30R_{\rm S}$ 
for given $\dot{m}_{\rm sup}$.
We divide the computational domain into 100 grids for the radial range and 160 grids for the angular range.
We employ the fixed grid size ratios of $\Delta r_{i+1}/\Delta r_i =1.05$ and $\Delta\theta_k /\Delta\theta_{k+1}=1.066$ in order to resolve the wind launching region. 
The number of grids is 134 for $z \lesssim 30 R_{\rm S}$ and $R\sim 100R_{\rm S}$ where the wind is vertically accelerated and the physical quantities drastically change (left side of the red dashed-dotted line in Fig \ref{fig4}).
The wind structure and the mass loss rate do not change so much
even if the spatial resolution near the launching region 
is lowered by employing $\Delta\theta_k /\Delta\theta_{k+1}=1.05$,
when the grid spacing in the $\theta$-direction 
becomes several times larger near the $\theta=\pi/2$ plane.
In addition, we note that the radiation force starts to 
accelerate the wind at the point slightly away 
from the numerical boundary of $\theta=\pi/2$ 
(see Fig. \ref{fig4}b). 
This means that the numerical resolution of our present simulations
is enough to resolve the launching of the wind.

Boundary and initial conditions are set by the same manner as N17.
We employ the axially symmetric boundary at the rotational axis of the
accretion disc ($\theta=0$). The outflow boundary conditions are applied
at the inner and outer boundaries (at $r_{\rm i}=30R_{\rm S}$ and
$r_{\rm o}=1500R_{\rm S}$), where
the matter can go out but is not allowed to go into the computational domain.
At the boundary of $\theta=\pi/2$, We apply the reflecting boundary condition.
At this plane, the radial velocity and the rotational velocity are fixed to be null and the Keplerian velocity. 
Also the temperature is fixed at the effective temperature of the standard disc model.
The density at $\theta=\pi/2$ plane is kept constant
at $\rho=\rho_0(M_{\rm BH},\varepsilon,r)$ that is the density at the surface of the standard disc, 
\begin{equation}
  \begin{array}{ll}
    &\rho_0(M_{\rm BH},\varepsilon,r)
    =\bar \rho(M_{\rm BH},\varepsilon,r)/(e-1)\\
    & =
    \left\{
    \begin{array}{l}
      5.24\times10^{-4}(M_{\rm BH}/M_{\odot})^{-1}(\varepsilon/\eta)^{-2}(r/R_{\rm S})^{3/2}\,{\rm g\,cm^{-3}}\\
      \qquad \qquad \qquad \qquad r\leq 18(M_{\rm BH}/M_{\odot})^{2/21}(\varepsilon/\eta)^{16/21}R_{\rm S} \\
      4.66(M_{\rm BH}/M_{\odot})^{-7/10}(\varepsilon/\eta)^{2/5}(r/R_{\rm S})^{-33/20}\,{\rm g\,cm^{-3}} \\
      \qquad \qquad \qquad \qquad r > 18(M_{\rm BH}/M_{\odot})^{2/21}(\varepsilon/\eta)^{16/21}R_{\rm S}\\
    \end{array} \right.
  \end{array}
  ,
\end{equation}
where ${\bar \rho} (M_{\rm BH},\varepsilon,r)$ is the vertically averaged density of the standard disc model \citep{SS73}.
We set $\varepsilon=\dot m_{\rm sup}$
at the region of $r\geq R_{\rm launch}$ and
$\varepsilon=\dot m_{\rm BH}=\dot M_{\rm BH}/\dot M_{\rm Edd}$ at $r< R_{\rm launch}$.

Here we note that the wind structure is insensitive to the density
profile at the boundary of $\theta=\pi/2$.
This is because that
the density at the boundary is too large for the line force to
accelerate the matter and the wind is accelerated from the point
slightly above the $\theta=\pi/2$ plane,
where the density is lower than $\rho_0(M_{\rm BH},\varepsilon, r)$ \citep[see also Appendix 2 in][]{N16}. This can be understood from the density dependence of the force multiplier in the low-ionization state region, $M\propto \rho^{-0.6}$.

We set the initial velocity to $v_r = v_\theta =0$ in the whole region.
The rotational velocity is set so as to meet the equilibrium between the gravitational force and the centrifugal force as 
$v_\varphi =(GM_{\rm BH}/r)^{1/2} \sin\theta$.
The initial temperature at a given point is set to $T(r,\theta)=T_{\rm eff}(r \sin\theta)$, where $T_{\rm eff}$ is the effective temperature of the standard disc. This means that  we have no temperature gradient in the vertical direction.
The initial density is given by assuming the hydrostatic equilibrium in the vertical direction as 
\begin{equation}
\rho (r,\theta)=\rho_0(M_{\rm BH},\varepsilon , r)
\exp \left ( -\frac{GM_{\rm BH}}{2c_{\rm s}^2 r \tan^2 \theta}\right),
\end{equation} 
where $c_{\rm s}$ is the sound velocity.

Radiation from the disc and X-ray source from 
within $r_{\rm i}$ are included but its wind is not
calculated under the assumption that these radii are too close to the
central X-ray source, so any potential wind material would be
overionized. 
This limitation makes 
the computational domain
small enough
that the calculation time is feasible. 
However, this is not entirely
justified, especially as our new disc structure reduces the
temperature of the inner disc emission below that predicted by a
standard disc. We will address this limitation in a future paper.

\section{Results}\label{sec:results}
\subsection{The effect of the wind on the mass accretion rate through
  the disc: a fiducial model}

\begin{figure}
  \includegraphics[width=\columnwidth]{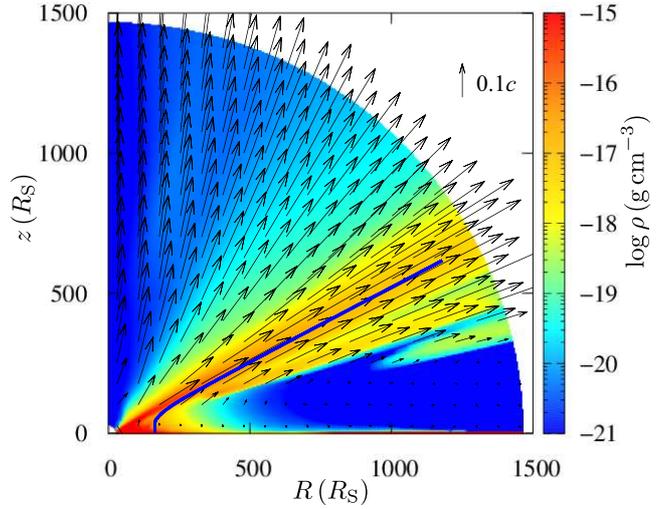}
  \caption{
    Time-averaged colour density map with the velocity vectors of the line-driven disc wind for 
    $M_{\rm BH} =10^8 \,M_\odot$, $\dot m_{\rm sup}=0.5$, and $f_{\rm X}=0.1$.
    The blue line shows a streamline through the mainstream
    (streamline passing through $\rho v_r$ peak at the outer boundary) of the wind.
    The $z=0$ plane roughly corresponds to the accretion disc surface and 
    the $z$-axis is the rotational axis of the disc.
  }
  \label{fig1}
\end{figure}

First we show the effect of the reduction in the mass accretion rate
through the inner disc in response to the mass loss rate of the
wind. We choose a fiducial set of parameters to illustrate this, with
$M_{\rm BH}=10^8 \,M_\odot$, $\dot m_{\rm sup}=0.5$ and $f_{\rm X}=0.1$. 
Fig. \ref{fig1} shows the
time averaged colour map of the wind density and velocity
structure 
in $R$-$z$ plane. Here, the $z$-axis is the rotational axis of the disc and $R$ is the distance from the $z$-axis. The $z=0$ plane corresponds to the $\theta=\pi/2$ plane.  
This shows many of the same qualitative features as seen in
the previous calculations in that there is a powerful funnel-shaped
wind, but there are also significant
differences in detail.
Without the 
reduction
in mass accretion rate through the inner disc, these initial
parameters produce a wind mass
outflow rate of $\dot{m}_{\rm out}=\dot{M}_{\rm out}/\dot{M}_{\rm Edd} \gtrsim 0.5$. This is clearly inconsistent as
the mass accretion rate supplied through the disc
$\dot{m}_{\rm sup}=0.5\lesssim \dot{m}_{\rm out}$.
Instead, 
in the present model (mass-conserved model),
the mass loss rate is reduced to  
$\dot{m}_{\rm out}\sim 0.23$, 
and the accretion rate through the inner disc and onto the black
hole is accordingly reduced to
$\dot{m}_{\rm BH}=\dot{m}_{\rm sup}-\dot{m}_{\rm out}\sim 0.26$ 
at a radius of $R_{\rm launch}\sim 59R_{\rm S}$.
Here, we note that a moderately fast wind 
in the polar region in Fig. \ref{fig1} would be driven by the gas
pressure force,
since the temperature of the matter in this region is very high, $\sim 10^9\,{\rm K}$.
Such a high temperature is probably induced by the shock heating.
The mass loss rate of this polar wind is quite small because of the low density.
Almost all of the mass ejected from the computational domain
is transferred by the line-driven funnel-shaped wind in our simulations.

\begin{figure}
  \includegraphics[width=\columnwidth]{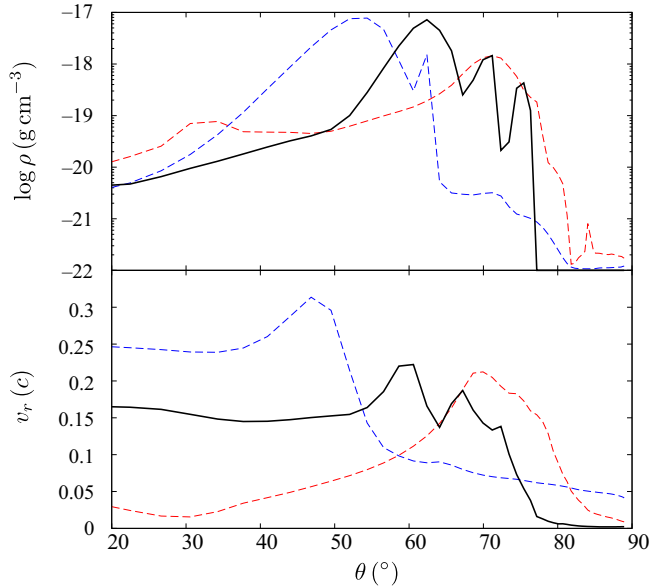}
  \caption{
    Angular profiles of the density (top panel) and the radial velocity (bottom panel) 
    at the outer boundary ($r=1500R_{\rm S}$).
    The black solid lines show the profiles for the new 
    mass-conserved simulations
    with $\dot m_{\rm sup}=0.5$ ($\dot m_{\rm BH}\sim 0.26$).
    The red dashed, and blue dashed lines show the profiles for the constant accretion rate models with 
    $\dot m_{\rm sup}=0.2$, and $\dot m_{\rm sup}=0.5$.
  }
  \label{fig2}
\end{figure}

Fig. \ref{fig2} shows the outflow properties at the outer boundary ($1500R_{\rm S}$) as a 
function of angle.
The black solid line shows the result of the new simulation, where the mass accretion rate
through the disc 
is reduced from $\dot m_{\rm sup}=0.5$ to $\dot m_{\rm BH}\sim 0.26$ at $R_{\rm launch}\sim 59R_{\rm S}$.
This is compared to previous results with constant mass accretion rate through the disc (dashed lines)
for $\dot m_{\rm sup}=0.5$ (blue) and $\dot m_{\rm sup}=0.2$ (red) that is close to the reduced $\dot m_{\rm BH}\sim 0.26$.
Both the density and the velocity of the wind in the new simulation peak at an angle which is intermediate between the two constant mass accretion rate simulations (see top and bottom panels). The peak value of the wind density is close to that of $\dot m_{\rm sup}=0.5$. In contrast, the wind maximum velocity is close to the result of $\dot m_{\rm sup}=0.2$. 

\begin{figure}
  \includegraphics[width=\columnwidth]{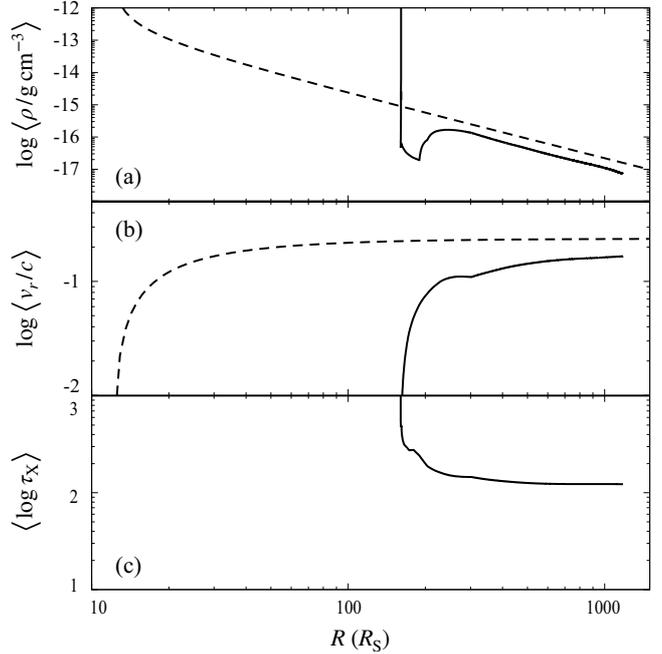}
  \caption{Density (panel a), velocity (panel b), and optical depth to the central X-ray source 
    (panel c) along the streamline.
    The angle brackets denote the time-averaged values.
    The horizontal axis is the distance from the rotational axis of the disc.
    The solid lines show the results along the mainstream (blue line in Fig. \ref{fig1})
    of our simulations for $M_{\rm BH} =10^8 \,M_\odot$, $\dot m_{\rm sup}=0.5$, and $f_{\rm X}=0.1$. 
    The dashed lines show the expectations from the phenomenological model of \citet{2015MNRAS.446..663H}
    with the wind terminal velocity of $v_\infty =0.237c$ and the mass loss rate of $8\,M_\odot {\rm yr^{-1}}$.
  }
  \label{fig3}
\end{figure}

\begin{figure}
  \includegraphics[width=\columnwidth]{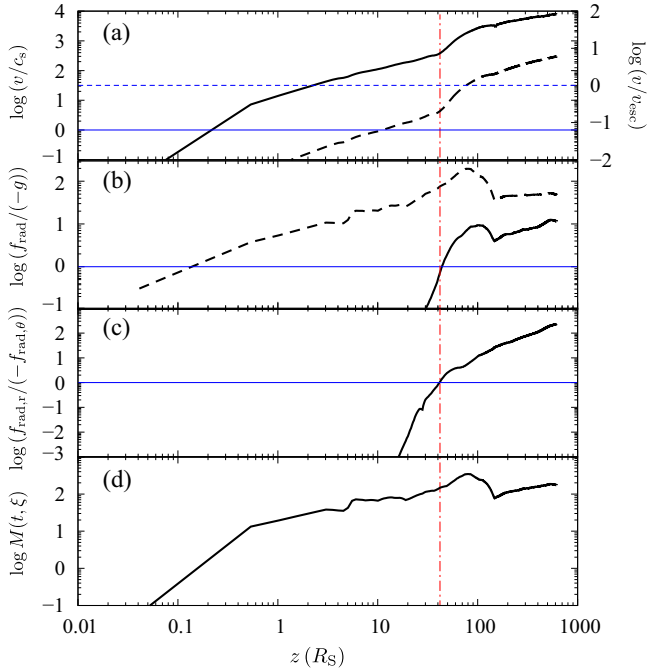}
  \caption{
    The velocity (panel a), 
    the radiation force normalized by the gravity (panel b),
    the ratio of the $r$-component of the radiation force to the
    $\theta$-component (panel c),
    and the force multiplier (panel d)
    on the major wind streamline (blue line in Fig 1).
    The horizontal axis is the distance from the $z=0$ plane.
    In the panel a, the velocity is normalized by the sound speed (solid line) and
    by the escape velocity (dashed line).
    The solid line in the panel b shows the $r$-component of the radiation force,
    and the $\theta$-component is presented by the dashed line.
    The wind matter mainly moves vertically
    at $z\lesssim 40R_{\rm S}$ (left side of red dashed-dotted line),
    and is blown away in the radial direction
    at $z\gtrsim 40R_{\rm S}$ (right side of red dashed-dotted line).
  }
  \label{fig4}
\end{figure}

The solid lines in Fig. \ref{fig3} show the time-averaged density,
velocity and optical depth along a streamline which goes
through the peak wind momentum loss at the outer boundary (the blue
line in Fig. \ref{fig1}).  This streamline originates on the disc at
$R\sim 160R_{\rm S}$, rises vertically upwards and then bends at
height of $\sim 40R_{\rm S}$ to become a radial line, so all the
vertical section of the streamline is at the same disc radius.  The
density (Fig. \ref{fig3}a) is high for the vertical section of the
streamline (at constant disc radius with $R\sim R_{\rm min}\sim
160R_{\rm S}$), and only drops to become $\propto R^{-2}$ when the
streamline bends.
Most of the acceleration takes place in this
vertical section, so the velocity increases rapidly at $R_{\rm min}$
(Fig. \ref{fig3}b, see also Fig. \ref{fig4}).
There is also a slower
acceleration zone where the wind velocity increases from $\sim\! 0.1c$
to $\sim \! 0.2c$
along the radial section of the streamline, from $R\gtrsim 160R_{\rm S}$.
This is a consequence of the ionization, which is very low at
these radii, with $\xi\sim 0$
so that UV line driving is very effective (see also Fig. \ref{fig4}).  This low
ionization parameter can be understood from the optical depth to
X-rays, which is always larger than $\tau_X\sim 100$
(Fig. \ref{fig3}c), so the X-rays cannot penetrate into this section
of the wind so the UV force multiplier is large.

We show details of the acceleration zone in our model in
Fig. \ref{fig4}, again along the main streamline shown by the blue
line in Fig. \ref{fig1} but this time plotted against vertical position
above the disc rather than radius.
The streamline rises vertically
from the disc, and bends radially at $z\sim 40R_{\rm S}$ (indicated by
the vertical red dashed-dotted line).  The velocity becomes supersonic near the
disc surface ($z\sim 0.2R_{\rm S}$, see the black solid line in
Fig. \ref{fig4}a compared to the horizontal solid blue line which marks $v=c_{\rm s}$) 
and becomes larger
than the escape velocity at $z\sim 70R_{\rm S}$ where the matter is 
accelerated toward $r$-direction (the black dashed line in
Fig. \ref{fig4}a compared to the horizontal blue dashed line which marks $v=v_{\rm esc}$). 
The dashed black
line in Fig. \ref{fig4}b shows the ratio of radiation force
to gravity in the $\theta$-direction. This exceeds
unity close to the surface of the disc ($z\sim 0.1R_{\rm S}$).  By
contrast, the solid black line shows this ratio in the $r$-direction, $f_{{\rm rad},r}/g_r$. This only starts to become
important around $z\sim 40 R_{\rm S}$.  These results mean that the
matter is first vertically accelerated and becomes supersonic due to
the radiation emitted from the disc right under the matter in the
region of $z\lesssim 40 R_{\rm S}$.
At $z\sim 40 R_{\rm S}$,
the radiation pressure from
the inner region also starts to become important.
This changes the
streamline trajectory to radial, but the material continues to accelerate 
as it still has substantial UV opacity so that 
the material reaches and exceeds the escape velocity.
We show the ratio of $f_{{\rm rad,} r}/f_{{\rm rad}, \theta}$ in
Fig. \ref{fig4}c.
This shows that the radial force becomes larger than the
$\theta$-component at $z\sim 40 R_{\rm S}$, changing the direction of
the streamline.
Fig. \ref{fig4}d indicates that the force multiplier
becomes $\sim\! 1$ at $z\sim 0.1R_{\rm S}$ and increases with the
distance from the disc surface ($z=0$ plane). This means that the line
force is comparable or larger than the radiation force due to the
electron scattering in $z\gtrsim 0.1R_{\rm S}$.  The main contributor
to the acceleration of the wind is the line force, even after $z\sim 40R_{\rm S}$
where the radiation from the central disc and X-ray
source are important.

However, this is a consequence of the assumption that the wind is
highly optically thick to X-ray radiation, with X-ray optical depth
$\tau_{\rm X}$ is equal to $100\tau_{\rm e}$ for $\xi\le 10^5$.  We
can see explicitly the effect of this assumption on the ionization
state by a comparison to the phenomenological biconical wind model of
\cite{2015MNRAS.446..663H} designed to fit to the ultra-fast outflow in
PDS456. This has an assumed (rather than calculated) velocity
structure such that $v_r(l)=v_0+(v_\infty -v_0)[1-R_{\rm min}/(R_{\rm min}+l)]$
where $v_\infty=0.237c$ (the dashed line in Fig. \ref{fig3}b), and an
assumed (rather than calculated) geometry where $R_{\rm min}=10$--$15R_{\rm S}$
is the launch radius of the wind.
Mass conservation then sets the
density $\propto (v_r R^2)^{-1}\to R^{-2}$ at large radii, and matches
quite well to our hydrodynamical model results
for $R\gtrsim 160R_{\rm S}$ (the dashed line in Fig. \ref{fig3}a).
However, this has very different ionization
structure with $\log\xi\sim 4$ 
which is much larger than the hydrodynamical results ($\xi\sim 0$).
This is because this model assumes that
$\tau_{\rm X}=\tau_{\rm e}$, so then the wind is never optically thick
along this sightline and the X-ray flux remains high.  We will explore
this further with more realistic photoionization cross-sections for
both the X-ray and line-driving fluxes in a future paper. Here we
simply recognize that $\tau_{\rm X}=100\tau_{\rm e}$ for $\xi\le 10^5$
and $\tau_{\rm X}=\tau_{\rm e}$ are opposite extreme assumptions, so we
calculate the wind properties using both of these to understand the
effect of photoionization on our wind.

\subsection{Wind mass loss rate as a function of X-ray illumination}
\label{sec:X-ray}
We now explore how the wind depends on the X-ray illumination.
In our method, we ignore the scattered and reprocessed X-ray photons and assume that the X-ray is emitted from the point source.
Although these assumptions are somewhat simpler to study the realistic dependence on the X-ray illiumination,
following parameter survey helps us to understand the effect of the X-ray on the wind properties.
Physically, we expect that the wind mass loss rate should depend on the strength of X-ray
illumination as the X-rays overionize the wind,
removing the UV opacity which is the driver for the wind launching
mechanism. This causes the wind to fall back to the disc (failed wind)
if it has not already reached its escape velocity at the point at
which it becomes ionized (PK04; \citealt{Risaliti10,Nomura13}).
The failed wind region creates a shadow which progressively
shields the proto-wind at larger radii from the ionizing flux, so that
eventually some material reaches the escape velocity and the wind is
formed at larger radii. This predicts some correlation of the wind properties
(increasing launch radius, decreasing velocity and decreasing mass
loss rate) with increasing central X-ray illumination.

The strength of X-ray illumination at any given point will also 
have some dependence on the geometry of the central X-ray source.
This is not well known at
present, so we assume that the X-rays are from a compact central
source, whose size is much smaller than
the inner radius of our computational domain, $r_{\rm i}=30R_{\rm S}$.
This is almost certainly not correct for the larger
values of $f_{\rm X}$, but we neglect this here so as to be able to
quantify the effect of a central X-ray source on the wind. 

\begin{figure}
  \includegraphics[width=\columnwidth]{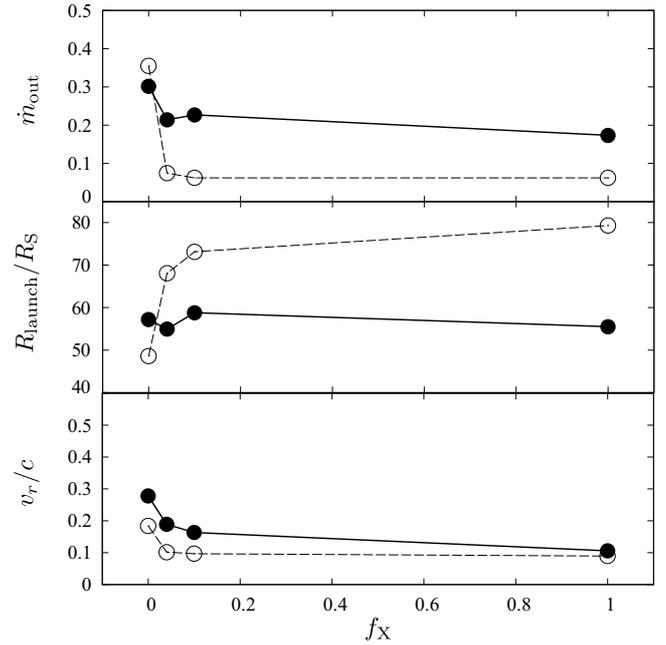}
  \caption{ Wind mass loss rate (top panel), launching radius (middle
    panel), and averaged radial velocity weighted by the density (bottom
    panel) as functions of X-ray fraction, $f_{\rm X}$.  We employ
    $M_{\rm BH} =10^8 \,M_\odot$ and $\dot m_{\rm sup}=0.5$ for $f_{\rm X}=0$, 0.04, 0.1 and 1.
    The filled circles show the
    results when we employ $\sigma_{\rm X}=\sigma_{\rm e}$ for $\xi >
    10^5$ and $100\sigma_{\rm X}$ for $\xi \leq 10^5$.  The open circles
    indicate the results using $\sigma_{\rm X}=\sigma_{\rm e}$
    independently of $\xi$ and ignoring the attenuation of the UV
    line-driving flux.
  }
  \label{fig5}
\end{figure}

The filled circles in Fig. \ref{fig5} show the wind outflow rates from
our fiducial model with $M_{\rm BH}=10^8\,M_\odot$ and $\dot{m}_{\rm sup}=0.5$ for $f_{\rm X}=0$,
0.04 and 1 as well as our fiducial
model with $f_{\rm X}=0.1$. There is indeed an anti-correlation of
mass loss rate with increasing $f_{\rm X}$, but the effect is rather
small, with only a factor 2 difference
between no X-ray illumination
($f_{\rm X}=0$)
and the model where the X-ray power is as strong as the total flux
from the disc ($f_{\rm X}=1$, top panel).
The wind launch radii are almost identical
for all models, in fact slightly decreasing for the higher $f_{\rm X}$,
contrary to the physical expectation that the requirement for
increasing shielding with increasing $f_{\rm X}$ would mean that the
wind is launched from further out (middle panel).  The averaged radial
wind velocity weighted by the density also shows the anti-correlation
with increasing $f_{\rm X}$, but the effect is small (bottom panel).

Hence even intense X-ray illumination does not destroy the wind for
this highly UV luminous disc \citep[see also Fig. 7 and 8 in][]{N16}.
However, the wind is strongly shielded from the central X-ray flux in
part due to the assumption that
$\tau_{\rm X}=100\tau_{\rm e}$ for $\xi\le 10^5$. By comparison,
the phenomenological biconical wind in \citet{2015MNRAS.446..663H} has $\xi\sim 10^4$
and its optical depth through the 
wind is $\tau_{\rm e}\sim 1$ at $76^\circ$ \citep[see Fig. 3 in][]{2015MNRAS.446..663H}.

We assess the impact of the assumption that
$\tau_{\rm X}=100\tau_{\rm  e}$ for $\xi\le 10^5$ by comparing this instead with $\tau_{\rm X}=\tau_{\rm e}$.
PK04 also explore this possibility, coupled with assuming that the UV optical depth is zero
rather than $\tau_{\rm UV}=\tau_e$. We follow their combination of
$\tau_{\rm X}$ and $\tau_{\rm UV}$ assumptions so we can compare results.
Fig. \ref{fig5} (open
circles) shows the wind properties with these different assumed X-ray
and UV optical depths.
The wind is similar for $f_{\rm X}=0$.
There is a steeper drop in the wind loss rate for
$f_{\rm X} =0$ to $0.04$ (top panel),
and a sharp increase in wind launch radius (middle panel)
as expected due to the wind being more ionized. However, there is
still a strong wind, even at $f_{\rm X}=1$, with mass loss rates across the
range of X-ray flux which are a factor $\sim\!2$--4 times smaller
than those of our fiducial model (filled circles). There is enough
shielding even for high X-ray luminosity since it only requires
$\tau_{\rm e}>1$ to reduce the X-ray ionizing flux substantially.
We will explore how robust our result is to a more detailed
photoionization model in a subsequent paper.

\subsection{Wind mass loss rate as a function of  $\dot{m}_{\rm sup}$ for realistic AGN SEDs}

\begin{figure}
  \includegraphics[width=\columnwidth]{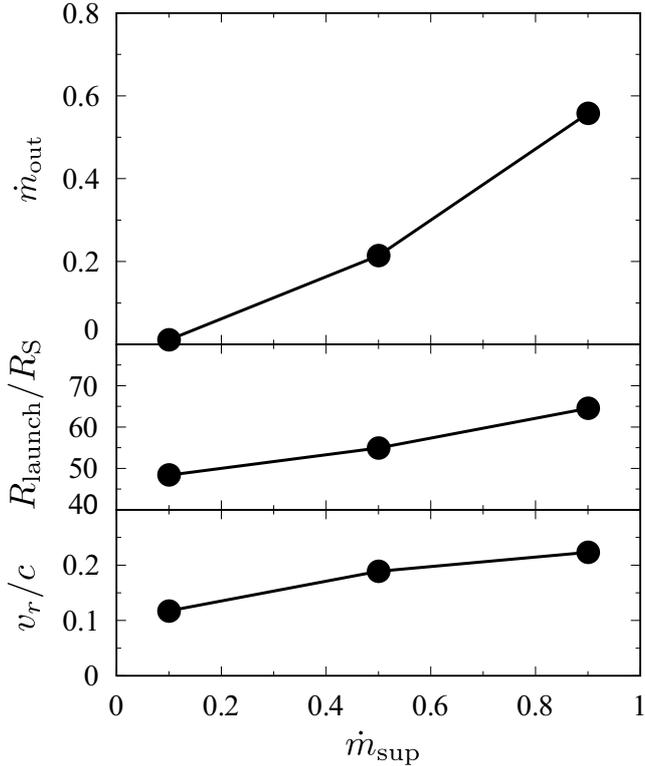}
  \caption{
    Wind mass loss rate (panel a), launching radius (panel b), and averaged radial velocity weighted by the density (bottom panel c) as functions the mass supply rate. 
    The black hole mass and the X-ray fraction are $M_{\rm BH} =10^8\,M_\odot$ 
    and $f_{\rm X}=0.2, 0.04, 0.02$ for $\dot{m}=0.1,0.5, 0.9$.
  }
  \label{fig6}
\end{figure}

Fig. \ref{fig6}a shows the effect of changing the black hole Eddington
fraction, for $\dot{m}_{\rm sup}=0.1, 0.5$ and $0.9$. We do this for
fixed $M_{\rm BH}=10^8\,M_\odot$ and $f_{\rm X}=0.1$, and also for a
probably more realistic AGN SED model where $f_{\rm X}$ decreases as
$\dot{m}_{\rm sup}$ increases
\citep{2012MNRAS.420.1848D,2012MNRAS.422.3268J}.
\citet{2018MNRAS.480.1247K} show that assuming $L_{\rm X}=f_{\rm X}L_{\rm D} = 0.02L_{\rm Edd}$
gives a fairly good match to the
observed data, so we also repeat the calculations for $f_{\rm X}=0.2$
for $\dot{m}_{\rm sup}=0.1$, 0.04 for $\dot{m}_{\rm sup}=0.5$ and
$0.02$ for $\dot{m}_{\rm sup}=0.9$.  However, as explained above, the
hard X-rays make very little difference to the wind properties in
these simulations so the results are very similar to those with fixed
$f_{\rm X}$ so we only show one set of results here.

As expected, the mass outflow rates in the wind increase for 
increasing $\dot{m}_{\rm sup}$,
with $\dot m_{\rm out} =1.1\times 10^{-2}$, $0.21$ and $0.56$ 
for $\dot m_{\rm sup}=0.1$, $0.5$, and $0.9$, which correspond to
$11\%$, $42\%$ and $62\%$ of the mass supply rates.
The wind efficiency
is a strong function of $\dot{m}_{\rm sup}$, and powerful UV line-driven
winds are not produced in our code below  $\dot{m}_{\rm sup}\sim 0.1$--$0.2$.
Fig. \ref{fig6}b shows the launch radius increases with $\dot{m}_{\rm sup}$, as
expected as the UV bright section of the disc is characterized by a constant temperature of $\sim\! 50,000\,$K
and this moves outwards with increasing $\dot{m}_{\rm sup}$.
The wind velocity slightly increases with $\dot{m}_{\rm sup}$ (Fig. \ref{fig6}c).
The outward shift of the UV zone caused by the large accretion rate tends to decrease
the wind velocity since the escape velocity decreases with an increase of the radius.
However, the larger mass accretion rate, the larger line-driving luminosity.
Thus, the wind velocity is thought to gradually rise with accretion rate.

\begin{figure}
  \includegraphics[width=\columnwidth]{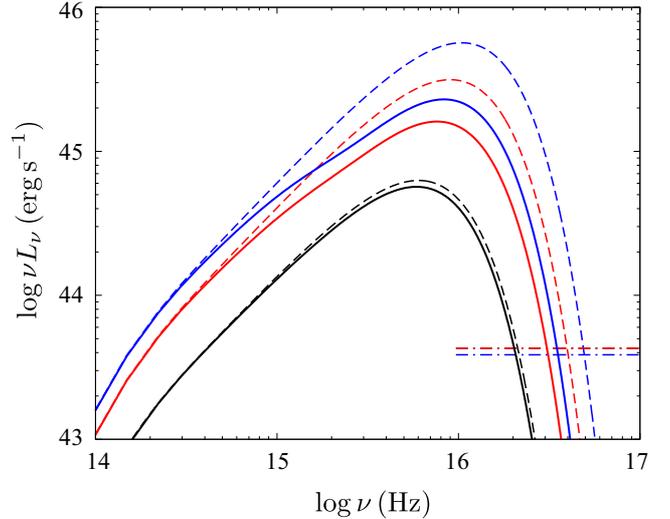}
  \caption{
    SEDs of the radiation from the accretion discs when we set $M_{\rm BH}=10^8 \,M_\odot$.
    The black, red, and blue solid lines show the SEDs for 
    $\dot m_{\rm sup} = 0.1$, $0.5$, and $0.9$ of our 
    mass-conserved simulations.
    The black, red, and blue dashed lines are the SEDs 
    of the standard disc model with $\dot m_{\rm sup}= 0.1$, $0.5$, and $0.9$.
    The black, red, and blue dot-dashed lines are the X-ray spectra 
    employing the power-law spectrum with a photon index of $\Gamma=2$.
    The X-ray fractions are $f_{\rm X}=0.2$, 0.04, and 0.02 for $\dot m_{\rm sup}=0.1$, 0.5, and 0.9.
  }
  \label{fig7}
\end{figure}

Fig. \ref{fig7} shows the resulting SED from the underlying disc emission for
the $10^8\,M_\odot$ case. The strong wind losses for $\dot{m}_{\rm sup}=0.9$
means that most of the accretion power is released as kinetic
luminosity rather than radiation, and the SED is much redder than
expected from a pure disc model. By contrast, the small wind losses at
$\dot{m}_{\rm sup}=0.1$ mean that the mass accretion rate through the disc
is not so different between the inner and outer regions, so it looks
almost like a standard disc. 

These spectra look very like the variable mass accretion rate disc spectra predicted by \citet{2014MNRAS.438.3024L}
which used O star winds to characterize the mass loss rate from each radius of the disc.
Their models assumed that there was no central X-ray source, so have maximal wind mass loss rates.
They reran their code for the specific case of $M_{\rm BH}=10^8\,M_\odot$ and $\dot{m}_{\rm sup}=0.1$ 
so we could directly compare. 
Their model had a powerful mass outflow, such that $\dot{m}_{\rm BH}\sim 0.02$ (0.05) if none (all) of the material lost in the wind
reaches the local escape velocity. However, this wind is launched at $15R_{\rm S}$, 
which is below our hydrodynamic inner grid limit of $30R_{\rm S}$. 
Thus our calculations cannot test whether such a wind is present. 

\section{Discussion}
\label{discussion}
\subsection{AGN feedback}
We present the first radiation hydrodynamic simulations of UV
line-driven disc winds which adjust the underlying disc emission to
compensate for the mass loss rates in the wind. This is extremely
important as previous calculations have inferred mass outflow rates
which are larger than the mass supply rate at the moderately high
Eddington ratios $\dot{m}_{\rm sup}\sim 0.5$.  The most powerful winds
in the local Universe are seen in similarly high $\dot{m}_{\rm sup}$
rates ($\sim\!0.5$ and above), e.g., PDS456 \citep{Reeves09} and
PG1211 \citep{Pounds03}, so these sources could not be well
modeled in previous work. 
These are the sources where 
the winds have enough power to affect star formation in the bulge of the host galaxy.
Our new results show that while the wind power is reduced from previous studies (N17), 
the winds are still powerful. We evaluate the 
kinetic luminosities of the winds by integrating over the outer boundary, so 
$L_{\rm kin}=2\pi r^2 \int^{89^\circ}_{0}\rho v_r^3 \sin \theta d\theta$ 
giving $\sim\! 4\%$, $\sim\! 25\%$, and $\sim\! 33\%$ of the disc luminosity for $\dot m_{\rm sup}=0.1$, $0.5$, and $0.9$ 
when the black hole mass is $M_{\rm BH}= 10^8 \,M_\odot$. 
These results are consistent with those estimated from the X-ray observations 
\citep{Tombesi12a,Gofford15}. 
 The model proposed by \citet{2005Natur.433..604D} suggests 
that the required kinetic luminosity for the feedback is at least 5\% of the bolometric luminosity, 
although \citet{Hopkins10} reduce the criteria down to 0.5\% 
based on the simulations focusing on a cloud in the interstellar medium exposed to the radiation and diffuse outflow from the AGN. 
\citet{2010ApJ...722..642O} performed large-scale one- and
two-dimensional calculations and found that winds can affect the
surrounding environment even if the kinetic luminosity is less than
$\sim 10^{-4}\dot M_{\rm acc}c^2$, which is not inconsistent with the estimation by \citet{Hopkins10}.

In addition to the mechanical feedback via the winds, the radiative
feedback would be also important 
for the large-scale inflow and outflow \citep[e.g.,][]{2009MNRAS.397.1791K}. 
Cosmological zoom-in simulations with the mechanical feedback via the winds and the radiative feedback 
using the simplified AGN feedback model 
are rapidly developed \citep[e.g.,][]{2017ApJ...844...31C,2018ApJ...860...14B}.
Our present model
can give the mechanical and radiative feedbacks quantitatively,
since the wind power and the disc luminosity are calculated 
as functions of the mass supply rate and the black hole mass.

\subsection{Limitations of current model and future works}
\label{sec:dis2}
There are several limitations of the current code, 
which are inherent in the current `standard' setup of hydrodynamics and radiation transfer for
UV line-driven disc wind simulations. 
We highlight the limitations of the radiative transfer, in particular the abrupt switch from
$\sigma_{\rm X}=\sigma_{\rm e}$ to $100\sigma_{\rm e}$ at $\xi < 10^5$ which means that the X-rays are heavily attenuated
below this ionization parameter.
This is a drastic overestimate of the opacity for material with $10^3\lesssim \xi\lesssim 10^5$,
which leads to the very efficient shielding.
Hence the effect of X-ray illumination on the wind structure is almost certainly underestimated in these calculations.
In the case of line-driving UV radiation, the opacity is assumed to be simply $\sigma_{\rm e}$,
which would underestimate the attenuation of the UV flux for $\xi\lesssim 10$.
This treatment is physically inconsistent with that the UV radiation is absorbed by the wind material
through the bound-bound transitions.
The model also does not include the ionizing effect of the disc photons above 200\,\AA, which will 
be substantial especially for the hotter discs predicted for smaller black hole masses and higher mass accretion rates. 
We will explore more physical models of X-ray and UV opacity in future work. 

Additionally, we ignore the scattered and reprocessed photons for simplicity in our method.
However, postprocessed radiation transfer calculations suggest that these secondary photons ionize the wind materials
\citep{Sim10, HP14}.
This might affect the line-driving mechanism and the resulting mass loss rate of the disc wind.
\citet{HP14} performed Monte Carlo simulation of the radiative transfer through the results of PK04.
They found that the ionization parameter estimated by Monte Carlo simulation including the secondary photons is
$\sim \! 4$ orders of magnitude larger than that from PK04 in the outflowing material,
though the ionization parameters are not so different in the high-density region ($\gtrsim \! 10^{14}\,{\rm g\,cm^{-3}}$).
The increase of the ionization parameter would weaken the line driving and reduce the mass outflow rate.
In order to quantitively assess the effect due to the secondary radiation,
the hydrodynamics simulations
coupled with the radiation transfer considering scattered and reprocessed photons are necessary.
Such simulations are not realistic due to the large computational costs at this time,
but the improvement of treatment of radiation transfer is an important future work in this field.

The computational domain
is fixed from $30$--$1500R_{\rm S}$, yet physically it would be better for
this to adapt in size to cover the disc UV temperature zone for the given black hole mass and mass accretion rate. 
The temperature in a standard disc is  $T^4\propto \dot{m}_{\rm sup}/M_{\rm BH}r^3$, and the UV zone 
extend inwards of $30R_{\rm S}$ for our lowest 
 $\dot{m}_{\rm sup}=0.1$ case. Including these inner radii 
would also enable us to test the \citet{2014MNRAS.438.3024L} UV line-driven disc wind models based on O star winds, 
and would allow us to calculate the wind mass loss rates for higher mass black holes, 
where the disc temperature is lower so the UV zone is at smaller radius and again extends below our fixed inner radius.
However, our disc does not have constant mass accretion rate, and the inner radii have lower temperature than predicted
due to the mass loss rate in the wind. At radii smaller than the wind launch radius the 
disc temperature is lower, $T^4\propto \dot{m}_{\rm BH}/M_{\rm BH}r^3$. Hence the
computational domain
should extend inwards of 
the initial predicted UV zone in order that it can properly capture the wind launch even after its temperature is adjusted for the mass loss rate. 
Just extending the grid inwards would mean that there are larger number of time-integration steps (caused by smaller grid spacing) and more grid points for the hydrodynamic calculation, which would slow the code down considerably. However, the outer disc in AGN should become self gravitating at a radius of only a few hundred $R_{\rm S}$, so shifting the same number of grid points inwards, which somewhat reduces computational costs, gives a much more physically realistic disc wind simulation.
However shifting the computational domain inwards means
that the 
X-ray illumination then depends on the unknown X-ray source geometry. 
Intriguingly, one potential source geometry is that the hard 
X-rays are powered by the mass accretion flow, so that the energy at small radii is dissipated in X-ray hot, optically thin plasma  rather than 
in the optically thick disc \citep{2012MNRAS.420.1848D,2018MNRAS.480.1247K}.
We will explore the effect of this and other potential source geometries. 

Our code considers the reduction of the mass accretion rate caused by the wind but does not solve the structure of the accretion disc itself. In order to obtain fully self-consistent results, it is necessary to perform the multidimensional simulations of the wind and the disc structure. In the current method, we assume that the geometrically thin and optically thick disc lies below the computational domain.
The boundary at $\theta=\pi/2$ corresponds to the disc surface. When we estimate the line-driving flux, the disc is treated as an external radiation source and the photons are supposed to be steadily emitted from the vicinity of the equatorial plane of the disc. Although the line force was not considered, global two-dimensional radiation hydrodynamics/magnetohydrodynamics simulations of the standard disc and the wind were performed by \citet{2006ApJ...640..923O}, \citet{2009PASJ...61L...7O} and \citet{2011ApJ...736....2O}.

Despite all these caveats, our model is the first hydrodynamic code 
to include the response of the disc to the mass loss rate in the 
line-driven wind. 
We show that the
resulting continuum spectra are different from those of the standard discs 
when the line-driven winds are launched (Fig. \ref{fig7}), similar to the numerical models of the disc structure 
based on O star wind mass loss rates by \citep{2014MNRAS.438.3024L}. 
The resulting spectrum should also be absorbed by the wind material along lines of sight which intersect the mass outflow,
so blueshifted absorption lines should be superimposed on the spectra. 
More detailed spectral models should include 
radiation transfer through this material, as have been done for the models of PK04
by \citet{Schurch09,Sim10,HP14}.

The time variation of the absorption lines \citep[e.g.,][]{Misawa07,Tombesi12b} is the remaining problem. 
This implies that the disc wind changes its structure in time and/or has 
non-axisymmetric clumpy structure. 
The density fluctuations of line-driven winds are found in one- or two-dimensional calculations \citep{OP99,Proga00}. 
In addition, recently, 
three-dimensional simulations of line-driven winds 
for cataclysmic variables reported the non-axisymmetric structure of the winds 
\citep{2018MNRAS.475.3786D,2018MNRAS.478.5006D}. 
Application of the three-dimensional calculations to AGNs 
and investigate the origin of the time variation are the important future works. 

Although we focus on the sub-Eddington sources in the present work, we need to investigate the super-Eddington cases in order to resolve the role of the outflow in the evolution of the most rapidly growing SMBHs. Radiation hydrodynamics simulations of super-Eddington sources revealed that radiation pressure on electrons accelerates the winds from the accretion discs 
\citep[e.g.,][]{2009PASJ...61L...7O,2011ApJ...736....2O,2013PASJ...65...88T,2018PASJ...70...22K}.
UV line driving would assist in launching disc winds as the luminosity approaches and then exceeds Eddington so that the radiation hydrodynamics simulations considering a combination of radiation pressure on electrons and the UV line driving are important for super-Eddington accretion flows. Such simulations are left as future work.

\section{Conclusions}
By performing the two-dimensional radiation hydrodynamics simulations, 
we found that the line-driven winds suppress the mass accretion onto the black hole 
especially in the luminous AGNs ($\dot m_{\rm sup} \gtrsim 0.5$).
Our simulations are the first 
mass-conserved
hydrodynamic models to include the 
reduction of the mass accretion rate through the inner disc 
due to the launching of disc winds. 

We show results for a  black hole mass of $M_{\rm BH}=10^8\, M_\odot$,
with mass supply rates of $\dot m_{\rm sup}=0.1$, $0.5$, and $0.9$. We find a powerful wind is generated in the 
latter two models, and the kinetic power of these winds is around 25\% of the disc luminosity, which is sufficient for AGN feedback
 \citep{2005Natur.433..604D,Hopkins10}.

The wind mass loss rate suppresses the mass accretion rate after the launching radius of the wind, producing a 
different accretion disc continuum spectra
which shows reduced flux in the UV
($\nu \gtrsim 10^{15}\,{\rm Hz}$) 
due to the large suppression of the mass accretion rate and 
corresponding low effective temperature in the inner disc.
For $\dot m_{\rm sup} \gtrsim 0.5$, 
there is clear difference between the SEDs predicted by our simulations 
and the SEDs assuming the constant accretion rate. 
This indicates that the line-driven wind imprints an observable feature into the continuum spectra, 
which may corresponds to a observed turnover in the UV at a constant temperature, 
modeled by \citet{2014MNRAS.438.3024L} using analytic/numerical calculations based on O star winds.

Our calculations here show the feasibility of producing a quantitative model for AGN feedback via UV line-driven winds.
We highlight a number of issues with the standard hydrodynamical disc wind setup which currently limit the reliability of these
models, 
but we will address these in a subsequent paper and apply the model to sources in the wide range of the black hole mass and the mass accretion rate. This would enable us to quantify the AGN feedback via winds across cosmic time.

\section*{Acknowledgements}
We would like to thank to Shane W. Davis for useful discussions. 
We also thank an anonymous referee for the constructive comments.
Numerical computations were carried out on Cray XC30 at the Center for Computational Astrophysics, 
National Astronomical Observatory of Japan. 
This work is supported in part by JSPS Grant-in-Aid for 
Scientific Research (A) (17H01102 KO), 
for Scientific Research (C) (16K05309 MN, KO; 18K03710 KO),
and for Scientific Research on Innovative Areas (18H04592 KO).
This research is also supported by MEXT as a priority issue (Elucidation of the fundamental laws and evolution of the universe) to be tackled by using post-K Computer and JICFuS.
CD acknowledges the Science and Technology Facilities Council (STFC)
through grant ST/P000541/1 for support, and  Kavli IPMU funding from the
National Science Foundation under Grant No. NSF PHY17-48958.




\appendix

\section{Comparison with previous works}
\subsection{Force multiplier}
\label{app:FM}
The force multiplier in our simulations
is consistent with the solution of
\citet[][hereafter CAK75]{1975ApJ...195..157C}
that is the one-dimensional line-driven stellar wind model.
Fig. 8 in CAK75 shows that the force multiplier is $\sim \!2$--5
near the sonic point.
Our result also indicates that the force multiplier is a few
at the sonic point (where $v=c_{\rm s}$, $z\sim 0.07R_{\rm S}$)
on the major wind streamline (see Fig. \ref{fig4}d).

Fig. \ref{fig4}d also shows 
the force multiplier in our model increases with the distances
and reaches $\sim\! 100$ at the point far away 
from the disc.
The increase of the force multiplier 
also occurs in CAK75, in which
the force multiplier 
goes up with an increase of the distances from the photosphere.
However, the force multiplier in the present model ($\sim\! 100$)
is much larger than that of CAK75 ($\sim\!$ a few).
The enhancement of the force multiplier is caused
by the reduction of the local optical depth parameter,
which is proportional to the density and inverse
proportional to the velocity gradient (see Eq. \ref{t-xi}).
The velocity gradient of our model is $\sim\! 10^{-4}c/R_{\rm S}$
--$10^{-3}c/R_{\rm S}$
since the wind is still slowly accelerated
at the point far away from the disc (Fig. \ref{fig3}b).
In contrast, the velocity gradient is very small in CAK75
since the wind almost reaches the terminal velocity.
In addition, the density
in our model decreases
with an increase of the distance more rapidly than in CAK75 model.
These density and velocity features cause the small local
optical depth parameter and the larger force multiplier.

The force multiplier continually increases for $R\gtrsim 100$,
but the total line force is the product of this
with the line-driving flux which roughly $\propto R^{-2}$.
Hence the total line force contributes most in the inner region
($R\sim 100 R_{\rm S}$ and $z\lesssim 30 R_{\rm S}$ 
for the major streamline).

Our simple method for evaluation of 
the velocity gradient ($dv/ds$)
does not lead to a large error in the force multiplier.
As we have mentioned in Section \ref{sec:line},
we evaluate $dv/ds$ along the direction of the line-driving flux
in order to suppress the computational cost.
However, strictly speaking, 
$dv/ds$ should be calculated along each light-ray.
Therefore, an ideal manner is calculating $dv/ds$ 
along the numerous light-rays from the disc surface
and evaluating the direction-dependent force multiplier.

In order to compare between our present method
and more accurate method,
we perform the test calculation
where we calculate $dv/ds$ along the $4096\times 4096$ light-rays
and estimate the force multiplier and line force
(This method is similar to that of PK04 but they prepare $12\times 12$ light-rays).
Here, we use the velocity distribution
of a snapshot of our simulations
for $M_{\rm BH}=10^8 M_\odot$ and $\dot m_{\rm sup}=0.5$.

As a result, we confirm that the line force estimated by our current method is quite similar to
that of the test calculation.
At the point on the major wind streamline,
where the matter is effectively accelerated
in the vertical direction
($R\sim 100R_{\rm S}$ and $z\sim 0.2R_{\rm S}$),
the $\theta$-component of the line force is
$f_{{\rm line},\theta} \sim -11.5\times 10^{-7}c^2/R_{\rm S}$
for our method and $\sim \! -8.6\times 10^{-7}c^2/R_{\rm S}$ 
for the test calculation.
The force multiplier in the present method,
$\sim\! 1.9$, is roughly consistent with
the $\theta$-component of the force multiplier,
$M_\theta=cf_{{\rm line},\theta}/\sigma_{\rm e} F_{{\rm line},\theta}\sim 1.2$, for the test calculation.
In both methods, 
the $r$-component of the line force is negligibly small 
compared to the $\theta$-component
because the $r$-component of the line-driving flux 
is quite small.

At the point where the outflow velocity exceeds the escape velocity
($R\sim 140R_{\rm S}$ and $z\sim 60R_{\rm S}$),
our present method gives 
$\left |f_{\rm line}\right | \sim 7.5\times 10^{-5}c^2/R_{\rm S}$, 
which is almost the same as the line force obtained 
by the test calculation, 
$\left |f_{\rm line}\right |  \sim\! 6.8\times 10^{-5}c^2/R_{\rm S}$. 
The force multiplier, $\sim\! 50$, is also quite similar 
to that obtained from the test calculation, 
$M_r=cf_{{\rm line},r}/\sigma_{\rm e} F_{{\rm line},r}\sim 44$ 
and $M_\theta\sim 45$.
These results show that our evaluation method for 
$dv/ds$ is reasonable.
More accurate results would be obtained by 
the radiation hydrodynamics simulations
in which the radiation transfer equations 
are solved along the many light-rays
\citep{2014ApJS..213....7J, 2016ApJ...818..162O}.

\subsection{Effective temperature of accretion disc and opacities}
Besides the method for estimation of $dv/ds$,
the effective temperature of the disc 
and opacities in our simulations 
are different from those of PK04.

In our method, the X-ray opacity is set to be $\sigma_{\rm X}=100\sigma_{\rm e}$ for $\xi \leq 10^5$ or $\sigma_{\rm X}=\sigma_{\rm e}$ for $\xi > 10^5$,
and the $r$-component of line-driving flux is attenuated
by supposing the opacity of $\sigma_{\rm UV}=\sigma_{\rm e}$
\citep[see also][]{Proga00}.
In contrast, $\sigma_{\rm X}$ is set to be $\sigma_{\rm e}$ independently of $\xi$ 
and $\sigma_{\rm UV}$ is assumed to be null in PK04.
As we explained in Sections \ref{sec:X-ray},
our treatment about $\sigma_{\rm X}$
leads the massive and fast wind (see Fig. \ref{fig5})
compared with the method of PK04.
Since neither our method nor method in PK04 is rigid, 
hydrodynamics simulations coupled with the frequency-dependent radiation transfer calculations 
are needed in order to obtain more realistic winds.
        
Next, we consider the impact of the setting of the effective temperature.
We use the simple radial profile of
\begin{equation}
  T_{\rm eff}\left( r \right)=T_{\rm in}\left( \frac{r}{3R_{\rm S}}\right)^{-\frac{3}{4}}.
\end{equation}
We set $T_{\rm in}$ to meet the condition of
\begin{equation}
  \dot m_{\rm sup}L_{\rm Edd}=\int_{\rm 3R_{\rm S}}^{r_{\rm out}}4\pi r\sigma T_{\rm eff}^4 dr,
\end{equation}
where we suppose
$T_{\rm eff}\left (r_{\rm out} \right )=3\times 10^3\, {\rm K}$
and $\sigma$ is the Stefan-Boltzmann coefficient.
In contrast, PK04 employ
\begin{equation}
  T_{\rm eff}\left (r\right)
  =T_{\rm in}\left( \frac{r}{3R_{\rm S}}\right)^{-\frac{3}{4}}\left( 1-\sqrt{\frac{3R_{\rm S}}{r}} \right)^\frac{1}{4},  
\end{equation}
where
\begin{equation}
  T_{\rm in}=\frac{3GM_{\rm BH}\dot M_{\rm sup}}{8\pi \sigma \left(3R_{\rm S}\right)^3}.
\end{equation}
Although we consider the disc emission from the regions of $3R_{\rm S} \lesssim r \lesssim r_{\rm out}$,
PK04 ignore the radiation from the inner part of the disc, $3R_{\rm S}\leq r < 30R_{\rm S}$,
where $T_{\rm eff}$ is $ \gtrsim 5\times 10^4\,{\rm K}$.

Here, we perform two test runs: (A) employing our current $T_{\rm eff}(r)$ and
(B) employing $T_{\rm eff}(r)$ described by PK04. 
In these tests, we employ $M_{\rm BH}=10^8\,M_\odot$, $f_{\rm X}=0.1$, and $\dot m_{\rm sup}=\dot m_{\rm BH}=0.5$.
Also, we set $\sigma_{\rm X}=\sigma_{\rm e}$ and $\sigma_{\rm UV}=0$ same as PK04.
The decreasing of the mass accretion due to the launching of the wind
is not taken into consideration in the test runs.
The results show that the mass loss rate estimated from the test run B is $\sim \! 4.9$ times larger than that obtained from the test run A.
The averaged radial velocity for the test run B ($v_r\sim 0.07c$),
which is consistent with that of PK04 ($v_r\sim\! 0.067c$),
is almost the same as that of test run A ($v_r\sim 0.096c$).

The difference of the mass loss rate is understood as follows.
The line-driving luminosity
(the luminosity integrated by the wavelength across the UV transition of 200--3200\,\AA,
see also Section \ref{sec:line})
emitted from the base of the wind ($r\geq 30R_{\rm S}$)
is $3.9\times 10^{44}\, {\rm erg\,s^{-1}}$ for the run A
and $1.9\times 10^{45}\, {\rm erg\,s^{-1}}$ for the run B.
Therefore, the matter near the disc surface 
would be effectively lifted up in the run B, leading to the larger mass loss rate.
Although the mass loss rate mainly depends on 
the line-driving luminosity of the wind base ($r\geq 30R_{\rm S}$),
the radial velocity of the wind would be determined by
the total line-driving luminosity.
For the run A, the total line-driving luminosity is 
$5.1\times 10^{45}\, {\rm erg\,s^{-1}}$.
This is comparable to the the line-driving luminosity 
for the run B, $1.9\times 10^{45}\, {\rm erg\,s^{-1}}$.
Note that the disc emission within $r < 30R_{\rm S}$
is neglected in the run B, so that 
the line-driving luminosity of the wind base 
equals to the total line-driving luminosity.
Since the efficiency of the acceleration in the radial direction of the wind
is thought to depend on the total line-driving luminosity,
the outflow velocity would become the same order in both runs.

To sum up,
our simple treatment for the calculation of $dv/ds$ does not affect the results.
Our assumption of $T_{\rm eff}\left( r \right)$ 
tends to suppress the wind mass loss but does not change the wind velocity.
The opacities employed in the present simulations drastically enhance the wind power.
Hence,
the reason why the wind velocity is higher for the present simulations than for the PK04
would be mainly caused by the difference of the opacities.
The detailed comparison among the various models is left as future work.
Numerical simulations employing realistic effective temperature profile 
with a detailed photoionization model are also important future works.


\bsp 
\label{lastpage}
\end{document}